\documentclass{moriond}

\def\be{\begin{equation}}
\def\ee{\end{equation}}
\def\bea{\begin{eqnarray}}
\def\eea{\end{eqnarray}}

\def\met{$E_T^{miss}$}
\def\lsp{{\tilde{\chi}_1^0}}
\def\sg{\tilde{g}}
\def\sq{\tilde{q}}

\newcommand{\Photo}{}

\begin{document}
\vspace*{4cm}
\title{INCLUSIVE SUSY SEARCHES AT THE LHC}

\author{ S. SEKMEN~\footnote{ssekmen@cern.ch} \\ on behalf of the ATLAS and CMS Collaborations}

\address{Physics Department, CERN, CH 1211 Geneva 23, Switzerland~\footnote{now at Physics Department, METU, Ankara, Turkey}}

\maketitle\abstracts{
I summarize the status of the inclusive SUSY searches conducted by the ATLAS and CMS experiments using the $\sim$20 fb$^{-1}$ of 8 TeV LHC data in the all inclusive, 0 lepton, $\ge 1$ lepton and $\ge 2$ lepton final states.  Current searches show that data are consistent with the SM.  The impact of this consistency was explored on a rich variety of SUSY scenarios and simplified models, examples of which I present here. 
}

The $\sim$20 fb$^{-1}$ of 8 TeV LHC data collected each by ATLAS~\cite{Aad:2008zzm} and CMS~\cite{Chatrchyan:2008aa} experiments have been explored thoroughly and deeply to catch a glimpse of new physics.  As a well-motivated theoretical framework with diverse realizations,  supersymmetry (SUSY) has been a favored guide while designing new physics searches.  However, the great variety of spectra, and production and decay channels requires us to think of every possibility and search everywhere.  Inclusive SUSY searches are designed to be sensitive to a wide range and variety of final states involving multiple objects such as jets, $b$-jets, leptons, photons and missing transverse energy.  Current analyses employ multiple search regions defined by different object multiplicities and different kinematic variables.  Having $\sim$20 fb$^{-1}$ data also allows us to work with disjoint search regions.  When SUSY is concerned, gluinos are the main target of inclusive searches due to their high production cross sections and diverse decay modes, but such searches are also sensitive to other sparticles, such as squarks and EW gauginos.  

Here, I will summarize the different types of ATLAS and CMS 8 TeV inclusive searches, by classifying them into final states of all inclusive, 0 lepton, $\ge 1$ lepton and $\ge 2$ lepton.  I will focus on searches and interpretations for gluinos and 1st/2nd generation squarks.  Dedicated searches for stops, sbottoms and direct production of EW gauginos are summarized in different contributions to this conference.

\begin{figure}[htbp]
\begin{center}
\includegraphics[height=4cm]{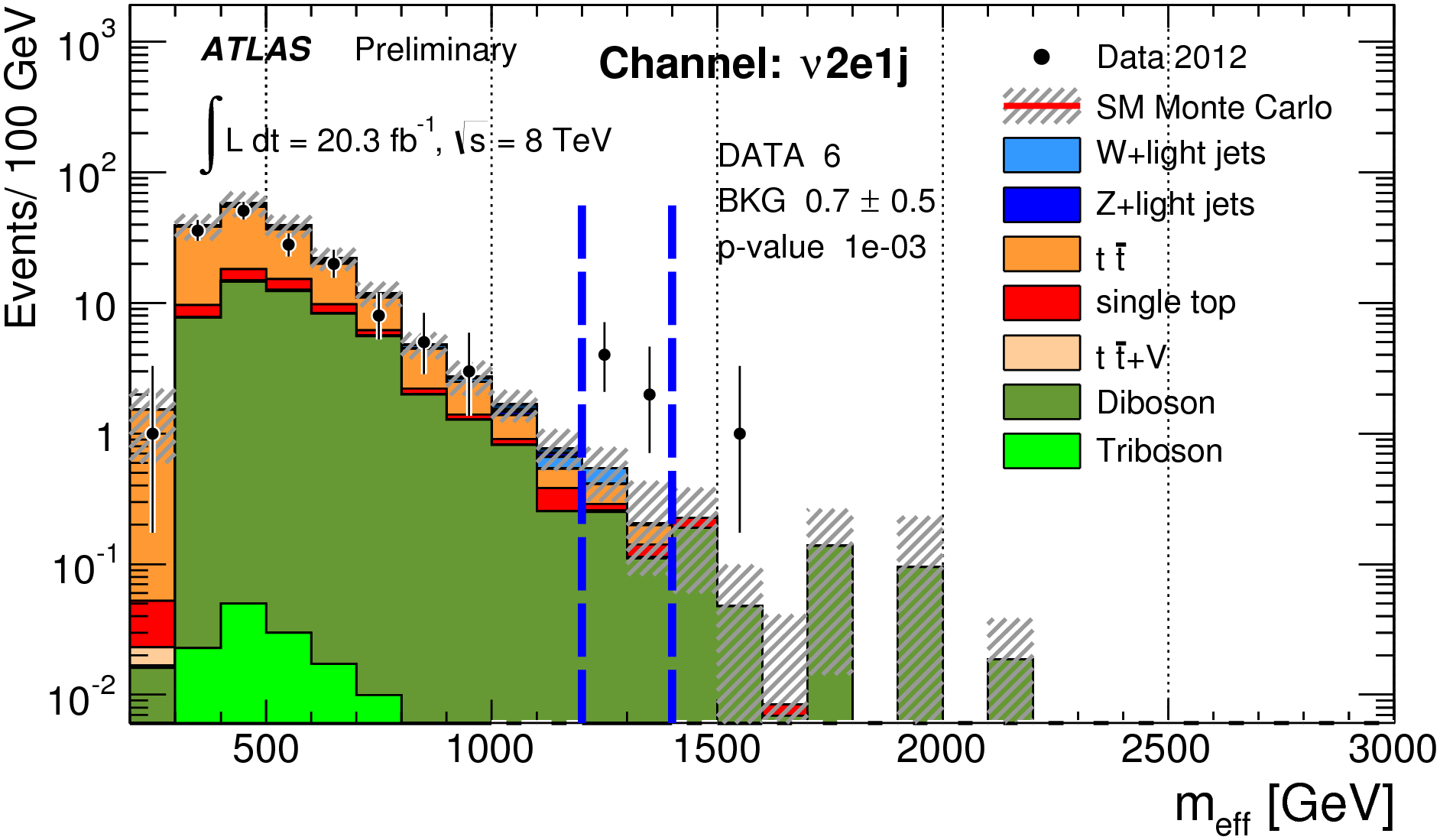} \qquad
\includegraphics[height=4cm]{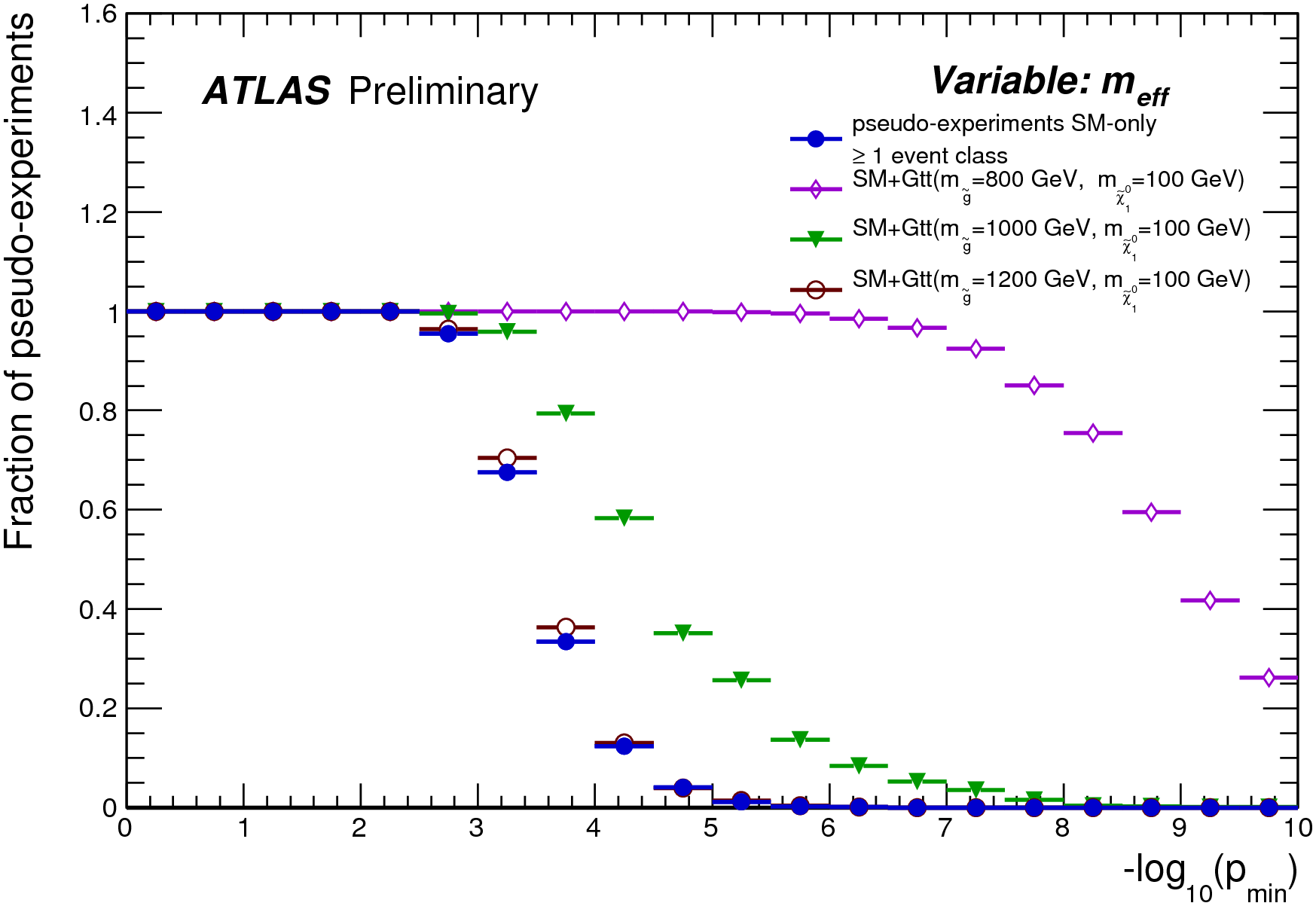} \\
\vspace{0.3cm}
\includegraphics[height=5cm]{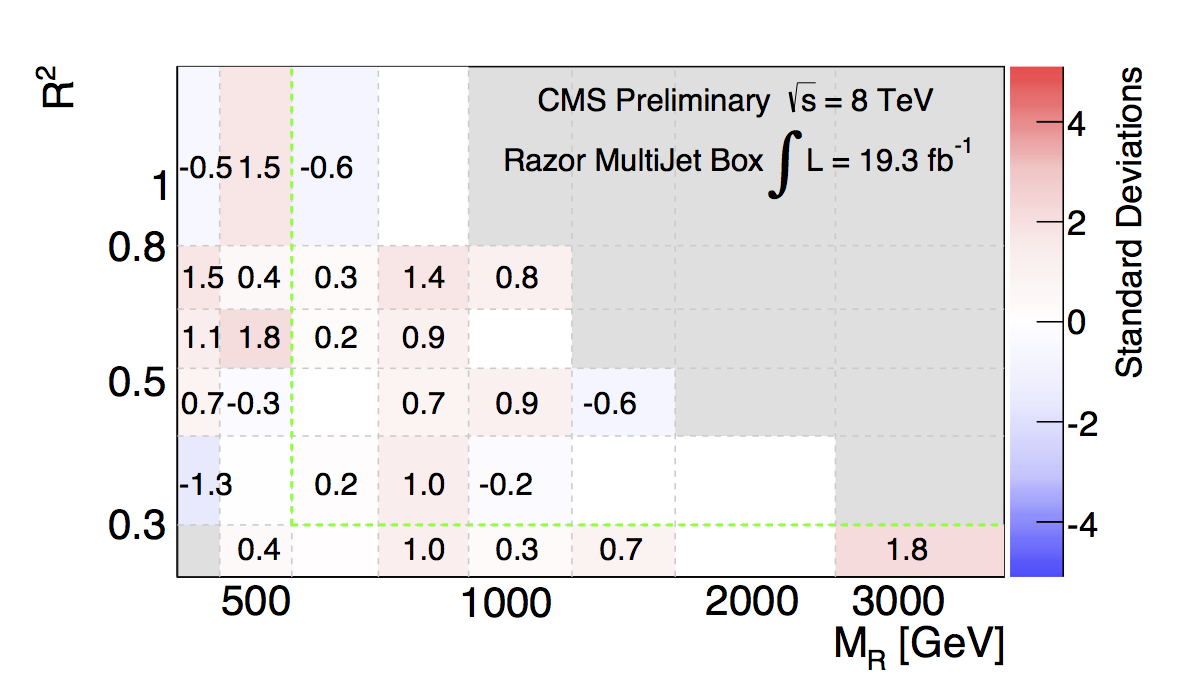} \qquad
\includegraphics[height=5cm]{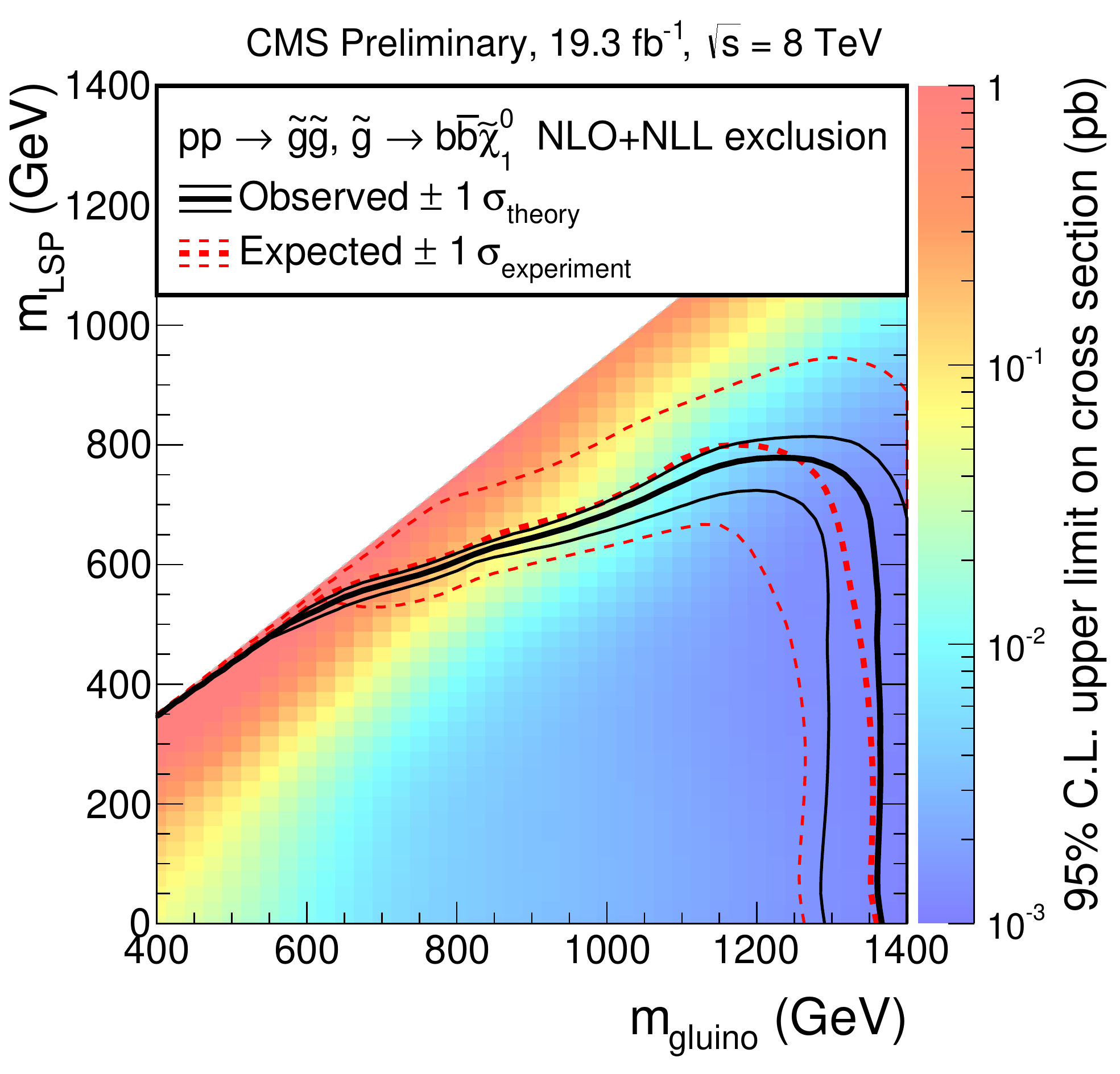} 
\caption{Top left: ATLAS general search: $m_{eff}$ distribution is shown for the event class with two electrons, one jet and \met . The dashed vertical lines indicate the region of interest which has the smallest p-value (0.0013) for this event class.  Top right: ATLAS general search: The expected fraction of pseudo-experiments having at least one event class with a  $-\log_{10}{\rm (p-value)}$ greater than the one shown on the axis for pseudo-experiments generated under the SM-only and SM$+$signal hypothesis for $m_{eff}$, where the signals are gluino models with $\sg \rightarrow t\bar{t}\lsp$.  The minimum $-\log_{10}{\rm (p-value)}$ for data for $m_{eff}$ is $\sim$3.  Bottom left: Comparison of the expected background and the observed yield in the MultiJet selection of the CMS razor analysis.  Bottom right: CMS razor razor limits on the $m_{\tilde{g}} - m_{\lsp}$ plane where $\tilde{g} \rightarrow b\bar{b} \tilde{\chi}_1^0$.
}
\label{fig:allinclusive}
\end{center}
\end{figure}

In a general search for new phenomena, ATLAS systematically classifies event topologies with isolated electrons, muons, photons, jets, b-jets and missing transverse energy~\cite{ATLAS-CONF-2014-006}.  697 event classes were scanned, and a deviation from SM was searched in distributions of effective mass, visible invariant mass and missing transverse energy.  Figure~\ref{fig:allinclusive} top left plot, $m_{eff}$ distribution is shown for the event class with two electrons, one jet and \met .  The analysis locates the regions of largest data-SM distribution in the kinematic distributions.  To test the consistency of the search with SM-only or SM$+$signal hypothesis, pseudo-data for event classes and kinematic distributions were generated for both hypotheses, and $p$ values were computed from the kinematic regions with largest deviation of the pseudo-data from the SM expectation.  Figure~\ref{fig:allinclusive} bottom right shows the expected fraction of pseudo-experiments having at least one event class with a  $-\log_{10}{\rm (p-value)}$ greater than the one shown on the axis for pseudo-experiments generated under the SM-only and SM$+$signal hypothesis for $m_{eff}$, where the signals are gluino models with $\sg \rightarrow t\bar{t}\lsp$.  
The minimum $-\log_{10}{\rm (p-value)}$ for data for $m_{eff}$ is $\sim$3, which means that data is consistent with the SM-only hypothesis, and the analysis would be sensitive to gluinos with mass smaller than $\sim$1 TeV.

On the other hand, CMS did an all inclusive search using the razor variables $M_R$ and $R^2$ which map the event into a dijet structure~\cite{CMS-SUS-13-004}.  Signal involving heavy particles is represented as a peak on the $M_R-R^2$ plane over a smoothly falling SM background distribution, which can be described using an analytical function.  This search compares data with SM background predictions on the $M_R-R^2$ plane for 9 event topologies involving jets, $b$-jets, electrons and muons.  Figure~\ref{fig:allinclusive} bottom left plot shows the discrepancy in standard deviations between data and SM prediction on the $M_R-R^2$ plane for the multijet topology.  In this topology and in all the rest, data is found to be consistent with SM, and hence limits were set on sparticle masses.  Figure~\ref{fig:allinclusive} bottom left plot shows the limits on the $m_{\tilde{g}} - m_{\lsp}$ plane where $\tilde{g} \rightarrow b\bar{b} \tilde{\chi}_1^0$.  

\begin{figure}[htbp]
\begin{center}
\includegraphics[height=5cm]{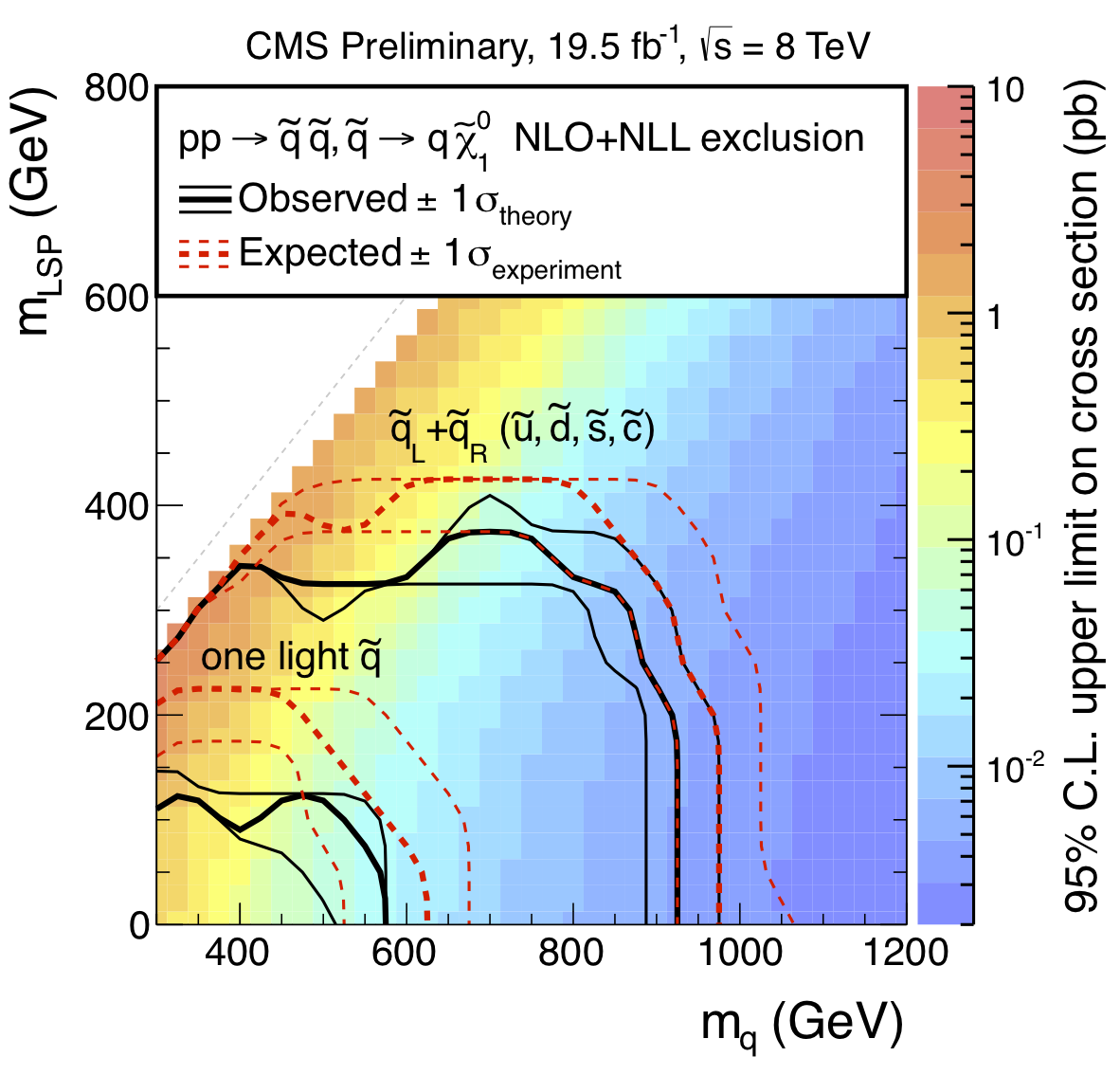} \qquad
\includegraphics[height=5cm]{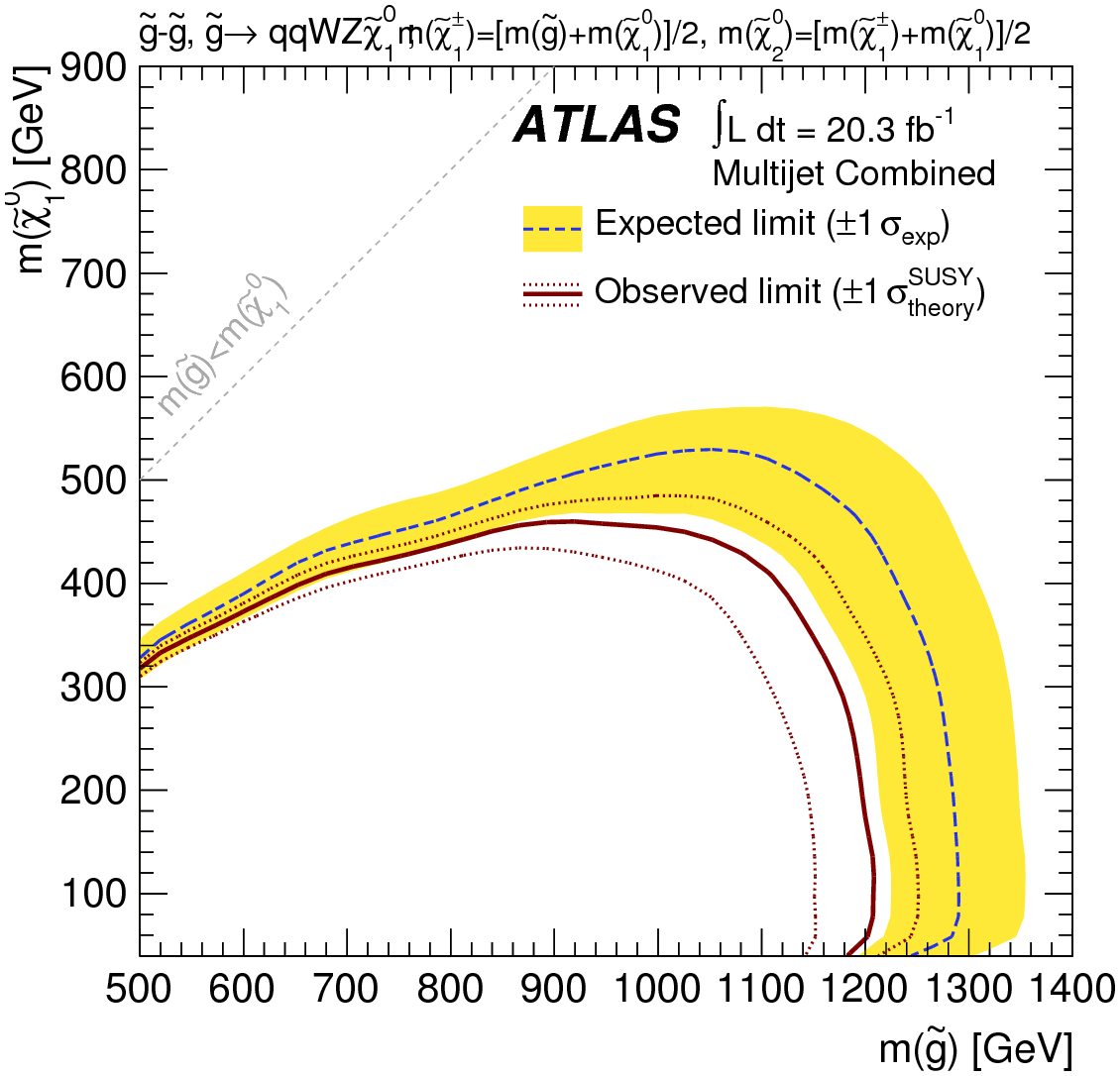}
\caption{Left: CMS $m_{T2}$ exclusion limits on the $m_{\sq} - m_{\lsp}$ plane for $\sq \rightarrow q\lsp$.  Right: ATLAS $\ge 7$ jet exclusion limits on the $m_{\sg} - m_{\lsp}$ plane for $\sg \rightarrow q\bar{q}\tilde{\chi}_1^\pm \rightarrow q\bar{q}W^\pm \tilde{\chi}_2^0 \rightarrow q\bar{q}W^\pm Z^0 \lsp$.}
\label{fig:0lepton}
\end{center}
\end{figure}

ATLAS and CMS have done a multitude of 0 lepton searches to look for sparticles in the hadronic decays using different kinematic variables~\cite{CMS-SUS-13-012}$^-$\cite{ATLAS-CONF-2013-061}.  An example is the CMS search that uses the stransverse mass variable $M_{T2}$ to discriminate signal~\cite{CMS-SUS-13-019}.  This analysis uses multiple search regions to target different sparticles.  For example, 0 $b$-jet bins are sensitive to 1st/2nd generation $\tilde{q}$s and $\sg$s, $b$-enriched regions with low jet multiplicity are aimed at $\tilde{t}/\tilde{b}$ production with decays to $b$s, while the high jet multiplicity regions are more sensitive to $\tilde{t}$ and $\tilde{b}$ production with decays to tops. Finally, the signal regions with $n_j \ge 3$ and $n_b \ge 3$ provide extra sensitivity to final states with multiple $\tilde{b}$s or $\tilde{t}$s, e.g. from gluino pair production.  Figure~\ref{fig:0lepton} left plot shows the impact of seing no signal in this analysis on the $m_{\sq} - m_{\lsp}$ plane for $\sq \rightarrow q\lsp$.  Another example is the ATLAS search with $\ge 7$ jets, which targets long decay chains from gluino production (and also some R-parity violating models).~\cite{ATLAS7jet}.  The sensitivity of the search is enhanced by considering the number of b-tagged jets and the scalar sum of masses of large-radius jets in an event, but in all cases data are consistent with the SM.  Figure~\ref{fig:0lepton} right plot shows the 95\% exclusion curve on the $m_{\sg} - m_{\lsp}$ plane for $\sg \rightarrow q\bar{q}\tilde{\chi}_1^\pm \rightarrow q\bar{q}W^\pm \tilde{\chi}_2^0 \rightarrow q\bar{q}W^\pm Z^0 \lsp$, where $m_{\tilde{\chi}^\pm_1} = (m_{\sq} + m_{\lsp})/2$ and $m_{\tilde{\chi}_2^0} = (m_{\tilde{\chi}_1^\pm} + m_{\tilde{\chi}_2^0})/2$.

\begin{figure}[htbp]
\begin{center}
\includegraphics[height=5cm]{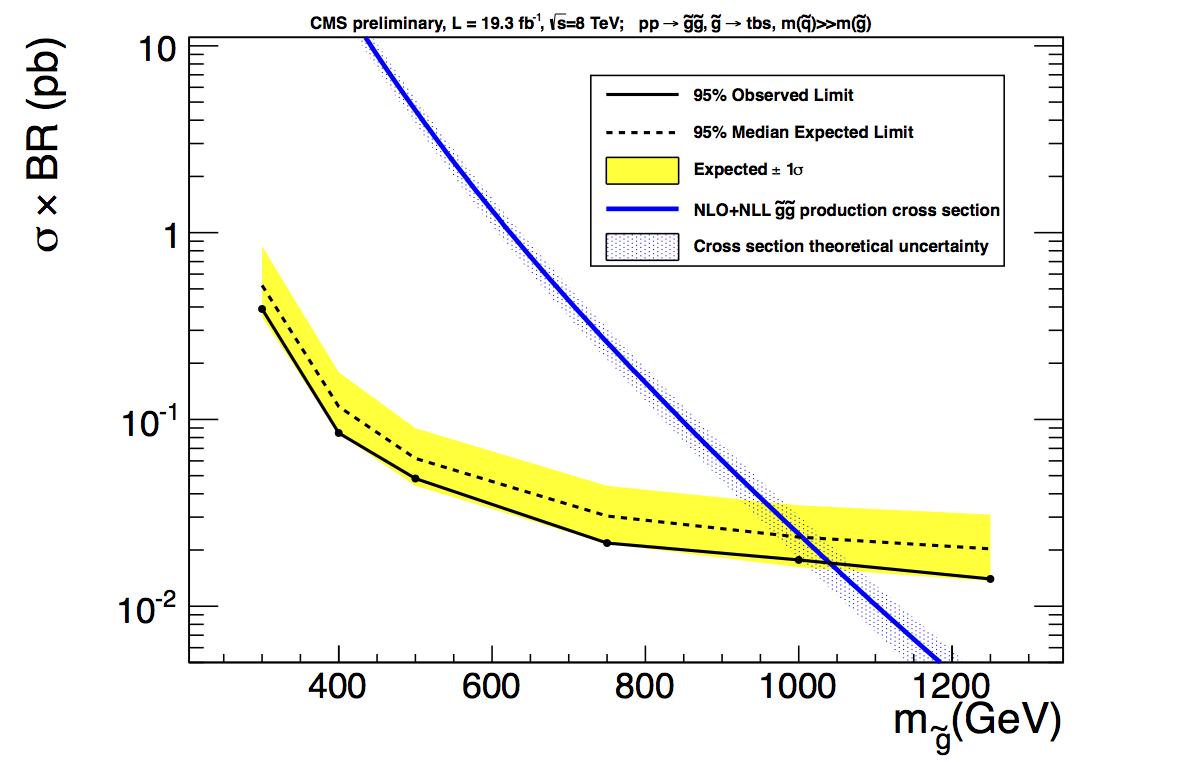}
\includegraphics[height=5cm]{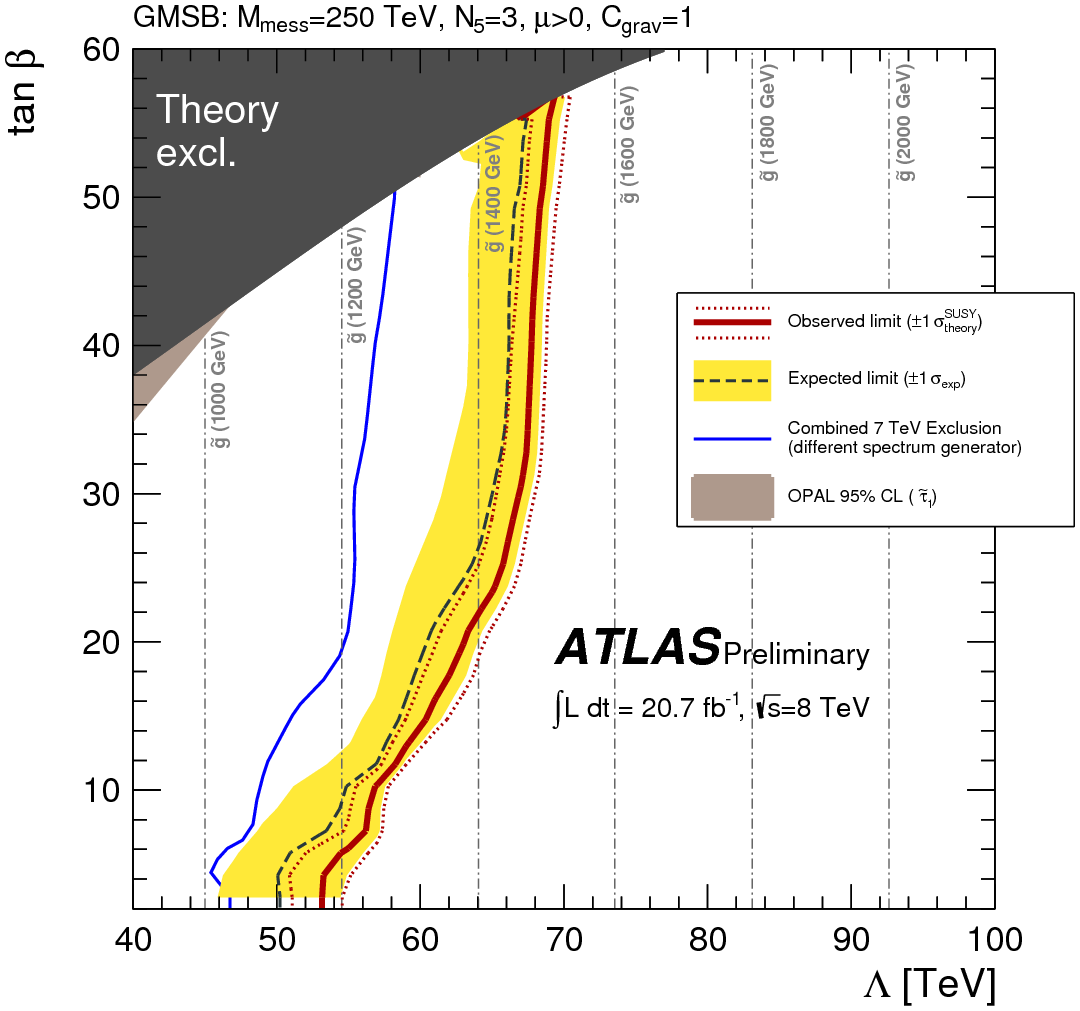}
\caption{Left: CMS .. exclusion limit on the gluino mass for R-parity violating SUSY where $\sg \rightarrow tbs$.  Right: ATLAS .. exclusion limit on the SUSY breaking mass scale in the low energy sector $\Lambda$ and ratio of Higgs VEVs $\tan\beta$ of the GMSB model.}
\label{fig:1lepton}
\end{center}
\end{figure}

Adding at least one lepton requirement to the searches~\cite{CMS-SUS-12-015}$^-$\cite{ATLAS-CONF-2013-026} enhances sensitivity to natural models.  One CMS search looks into signatures with $\ge 1$ leptons, $\ge 6$ jets and no \met ~\cite{CMS-SUS-12-015}.  To separate signal from the SM backgrounds, the analysis uses $b$-jet multiplicity.  The SM backgrounds were predicted by correcting MC to match the $b$-tag efficiency and mistag rate measured in data.  Having seen no excess in data, limits were set on models with R-parity violation and minimal flavor violation, in particular with pair-produced gluinos decaying each to a top, a bottom, and a strange quark.  Figure~\ref{fig:1lepton} shows the 95\% CL exclusion limit on the gluino mass for the R-parity violating case.  On the other hand, the ATLAS search~\cite{ATLAS-CONF-2013-026} concentrates on final states with $\ge 1 \tau$, jets, \met , and no extra light leptons.  This search is particularly sensitive to models with $\tau$-rich signatures such as the stau-coannihilation case of the constrained MSSM, GMSB with $\tilde{\tau}$ as the next to lightest SUSY particle (NLSP), and natural gauge madiation models with stau NLSP.  Figure~\ref{fig:1lepton} right plot shows exclusion limits set by this analysis on SUSY breaking mass scale in the low energy sector $\Lambda$ and ratio of Higgs VEVs $\tan\beta$ of the GMSB model for messenger mass $M_{mess} = 250$~TeV, number of messenger fields $N_5 = 3$, higgsino mass parameter $\mu > 0$ and gravitino mass scale factor $C_{grav} = 1$.

\begin{figure}[htbp]
\begin{center}
\includegraphics[height=5cm]{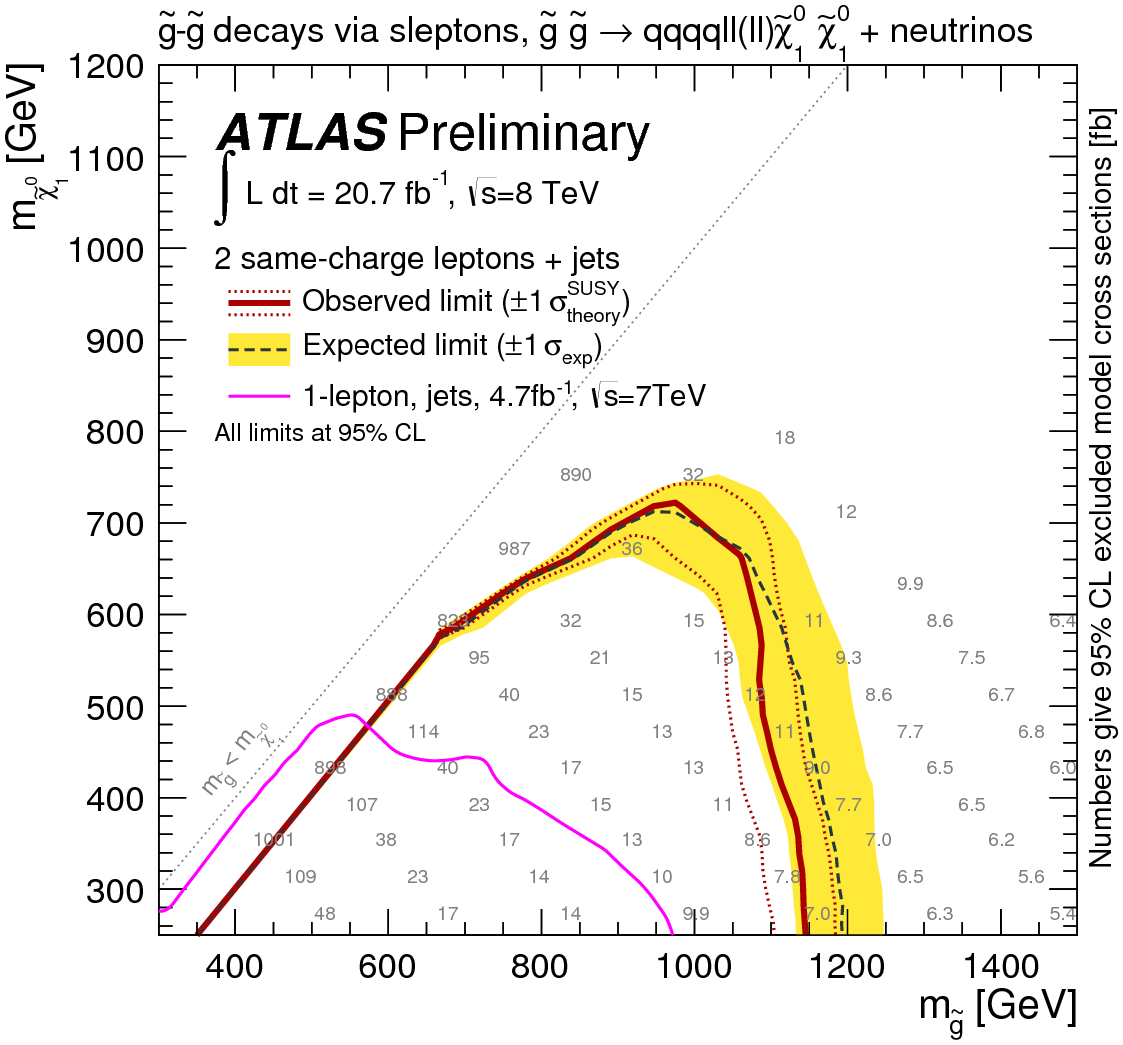} \qquad
\includegraphics[height=5cm]{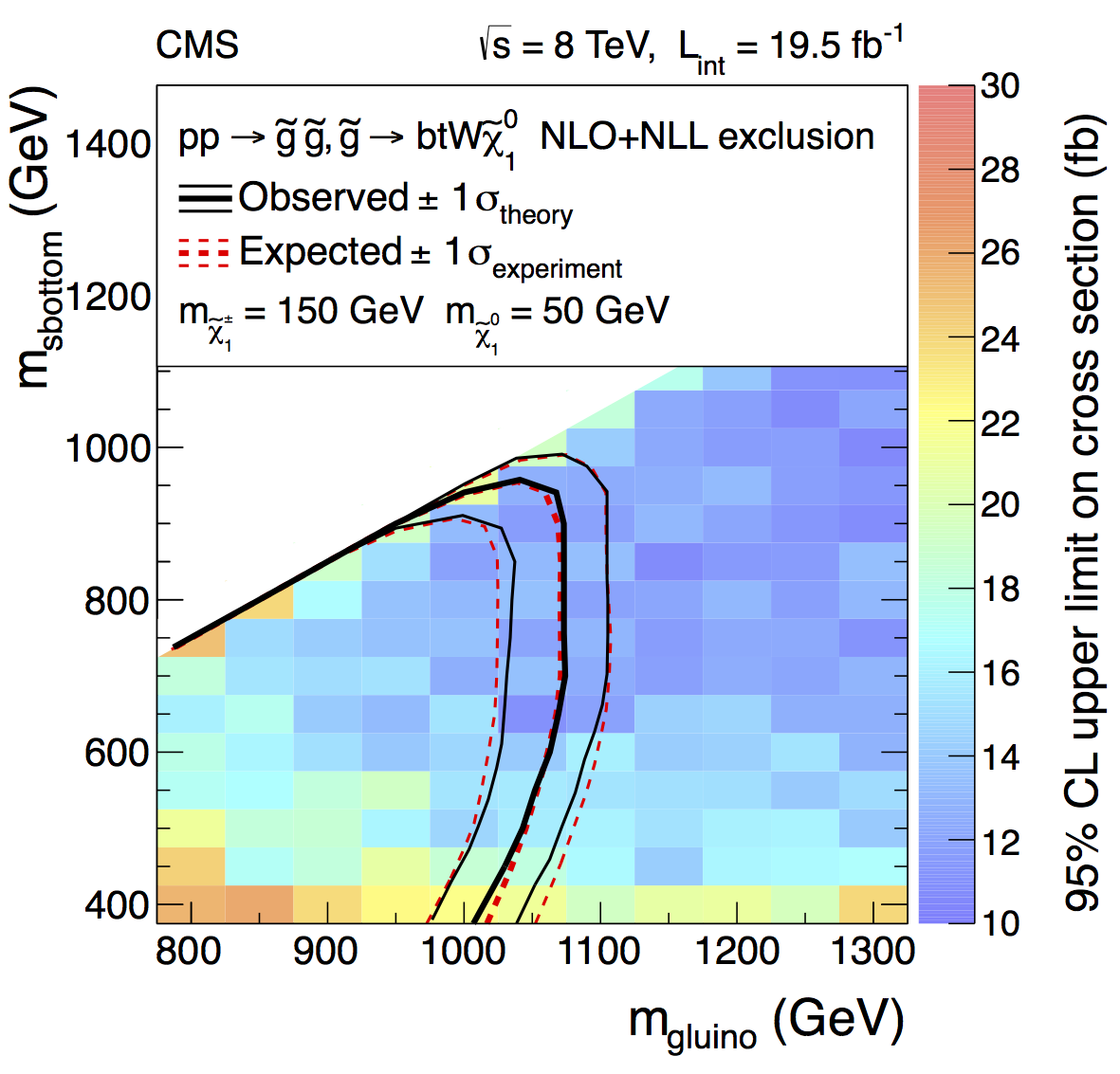}
\caption{Left: ATLAS same-sign dilepton exclusion limits on the $m_{\sg} - m_{\lsp}$ plane for gluinos that follow a decay chain via squarks, charginos or neutralinos, and sleptons or sneutrinos.  Right: CMS same-sign dilepton exclusion limits on the $m_{\sg} - m_{\lsp}$ plane for $\sg \rightarrow btW\lsp$.}
\label{fig:2lepton}
\end{center}
\end{figure}

Then there are searches involving two leptons~\cite{ATLAS-CONF-2013-089}$^-$\cite{CMS-SUS-13-016}, which are particularly sensitive to gluino decays involving stops, sleptons and EW gauginos.  The ATLAS search~\cite{ATLAS-CONF-2013-007} looks into final states with jets, $b$-jets, \met  and two leading leptons ($e$ or $\mu$) with $p_T > 20$ and same electric charge.  Three signal regions are defined with $0$, $\ge 1$ and $\ge 3$ $b$-jets and cuts on $n_{jets}$, \met, transverse mass $m_T$ and effective mass.  Again, data is consistent with the SM, and limits are set on constrained MSSM and a wide variety of simplified models.  Figure~\ref{fig:2lepton} left plot shows the exclusion limit on the $m_{\sg} - m_{\lsp}$ plane for gluinos that follow a decay chain via squarks, charginos or neutralinos, and sleptons or sneutrinos.  A corresponding CMS same sign dilepton analysis uses the hadronic transverse momentum variable $H_T$ to discriminate signal, and studies high and low lepton $p_T$ regions to target different SUSY scenarios, and also looks at low \met regions to target RPV SUSY ~\cite{CMS-SUS-13-013}.  Figure~\ref{fig:2lepton} right plot shows the impact of seing no signal in this search on the $m_{\sg} - m_{\lsp}$ plane for $\sg \rightarrow btW\lsp$.

\begin{figure}[htbp]
\begin{center}
\includegraphics[height=5cm]{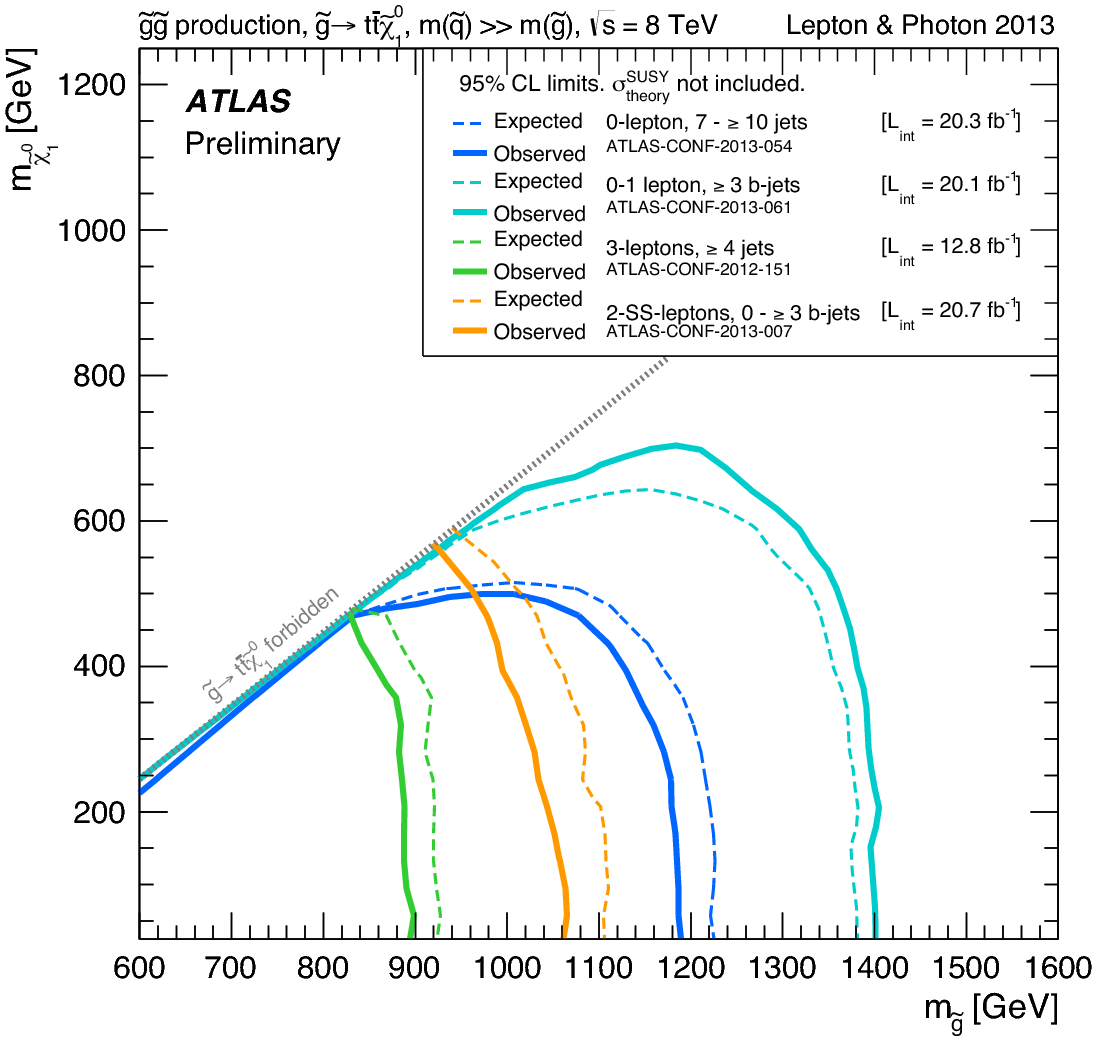}
\includegraphics[height=5cm]{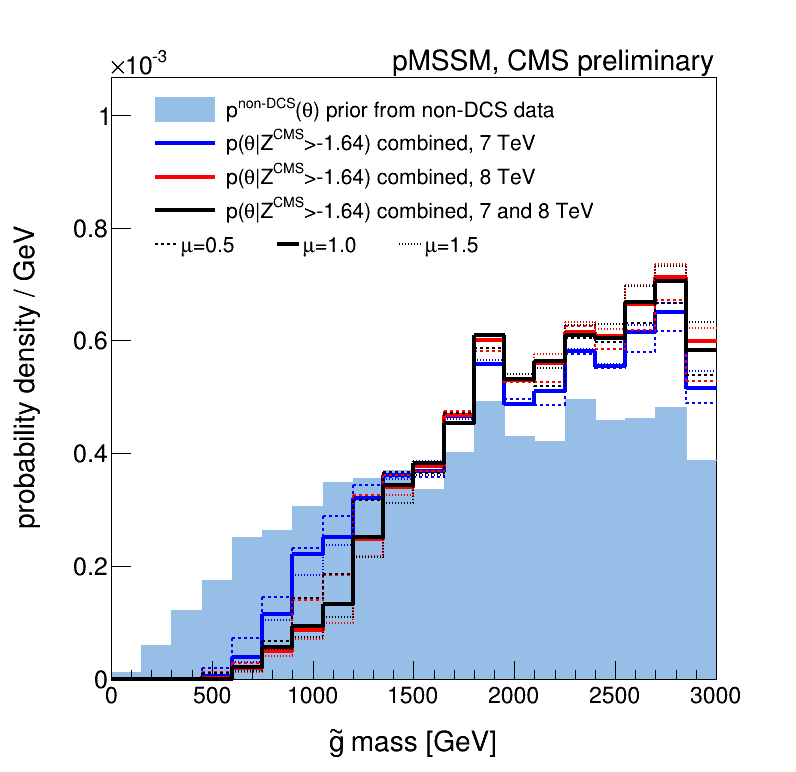}
\includegraphics[height=5cm]{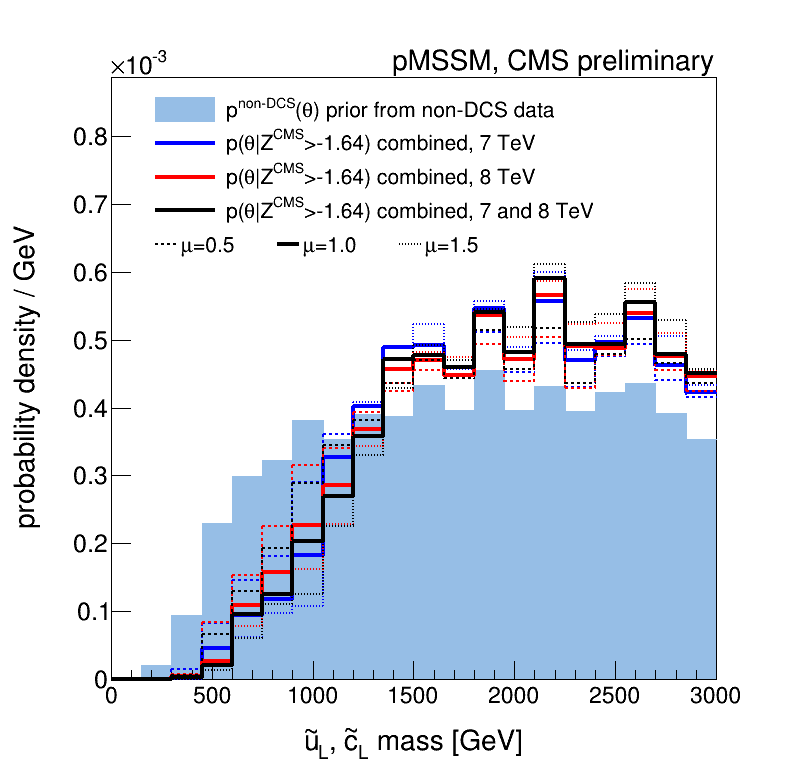}
\caption{Left: Impact of various inclusive ATLAS searches on the $m_{\sg} - m_{\lsp}$ plane for $\sg \rightarrow t\bar{t}\lsp$.  Middle and right: Probability distributions of $\sg$ and $\tilde{u}_L, \tilde{c}_L$ squark masses in pMSSM.  Filled blue histograms show the effect of various $b$ physics, EW, Higgs, top measurements, and line histograms show the impact of the combination of various 7 TeV, 8 TeV and 7+8 TeV inclusive CMS searches. }
\label{fig:interpretation}
\end{center}
\end{figure}

The impact of seing no signal on all these inclusive searches have been investigated in many more SUSY models and simplified models.  Overall, inclusive searches can explore gluinos up to mass $\sim 1.4$~TeV.  For example, Figure~\ref{fig:interpretation} left plot shows the summary of exclusions from various ATLAS searces on the $m_{\sg} - m_{\lsp}$ plane for $\sg \rightarrow t\bar{t}\lsp$.  ATLAS has also interpreted searches extensively using constrained MSSM.  CMS, on the other hand, has done a thorough interpretation on the p(henomenological) MSSM model, which is a 19-dimensional realization of SUSY at the SUSY scale.  A global Bayesian analysis was performed on the pMSSM space to obtain posterior probability densities for model parameters, masses and observables after the combined impact of $b$ physics, EW, Higgs, top measurements and various 7+8 TeV CMS SUSY searches.  Middle and right plots in Figure~~\ref{fig:interpretation} show that CMS inclusive searches do have a visible impact on gluino and $\tilde{u}_L, \tilde{c}_L$ squark mass probability distributions.  ATLAS and CMS are currently getting ready for the next phase of LHC data taking with 13-14 TeV energy, which we hope, may bring us a hint of new physics.  We must note that the 8 TeV results help us characterize the yet unexplored regions and design more efficient analyses for the pursuit of new physics.

\section*{Acknowledgments}

I would like to thank my colleagues in ATLAS and CMS Collaborations for their hard work in producing the results in this note, and to the organizers of Moriond QCD 2014 for a fruitful conference.

\end{document}